\documentclass{ws-procs9x6}

\def\beq{\begin{equation}}
\def\eeq{\end{equation}}
\def\bea{\begin{eqnarray}}
\def\eea{\end{eqnarray}}

\def\cal{\mathcal }  

\def\beq{\begin{equation}}
\def\eeq{\end{equation}}
\def\bea{\begin{array}}
\def\eea{\end{array}}

\def\del{\partial }
\def\to{\rightarrow}

\def\[{\left[}
\def\]{\right]}
\def\({\left(}
\def\){\right)}

\def\sm0{{\widetilde{m}_0}}

\def\ov{\overline}

\def\U1em{{U(1)_{\rm em}}}
\def\to{\rightarrow}

\def\sq2{\sqrt{2}}

\def\tt{t\bar{t}}

\def\End{\end{document}}
\def\ra{\rightarrow}

\def\Journal#1#2#3#4{{#1} {\bf #2} (#4) #3}

\def\PLB{{\em Phys. Lett.}  B}

\begin{document}
\title{
Testing Supersymmetry in $A H^\pm$ Associated Production
\footnote{\uppercase{T}alk presented by 
\uppercase{C.P.Y.} at {\it \uppercase{SUSY} 2003:
\uppercase{S}upersymmetry in the \uppercase{D}esert}\/, 
held at the \uppercase{U}niversity of \uppercase{A}rizona,
\uppercase{T}ucson, \uppercase{AZ}, \uppercase{J}une 5-10, 2003.
}
}

\author{Qing-Hong Cao$^1$, Shinya Kanemura$^2$  and  C.--P. Yuan$^1$}

\address{
$^1$Department of Physics and Astronomy, Michigan State University,
     East Lansing, Michigan 48824-1116, USA \\
$^2$Department of Physics, Osaka University,
         Toyonaka, Osaka 560-0043, Japan \\
E-mail: cao@pa.msu.edu, kanemu@het.phys.sci.osaka-u.ac.jp, yuan@pa.msu.edu}


\maketitle

\abstracts{
In the Minimal Supersymmetric Standard Model, the masses of the
charged Higgs boson ($H^\pm$) and the CP-odd scalar ($A$) are
related by $M_{H^+}^2=M_A^2+m_W^2$ at the Born level.
Because the coupling of $W^-$-$A$-$H^+$ is fixed by gauge interaction,
the Born level production rate of
$q \bar q' \to W^{\pm \ast} \to A H^\pm$ depends only on
one supersymmetry parameter -- the mass ($M_A^{}$) of $A$.
We illustrate how to test the mass relation
between $A$ and $H^+$ in case that the signal is found at the LHC.
If the signal is not found, the product of the decay branching
ratios of $A$ and $H^\pm$ predicted by the MSSM
is bounded from above as a function of $M_A$.
}

One of the top priorities of current and future high-energy
colliders, such as the Fermilab Tevatron and CERN Large Hadron
Collider (LHC), is to probe the mechanism of the electroweak
symmetry breaking. In the Standard Model (SM) of particle physics,
this amounts to searching for the yet-to-be-found Higgs boson. It
is also possible that the mechanism of electroweak symmetry
breaking originates from new physics beyond the SM. 
Supersymmetry (SUSY) is one of the most commonly studied new 
physics models. The Higgs sector of the Minimal Supersymmetric
Standard Model (MSSM) is known as a special case of the Two Higgs
Doublet Model (THDM) with the type-II Yukawa interaction~\cite{hhg},
and contains five physical scalar
states, i.e., two CP-even ($h$ nd $H$), a CP-odd ($A$) and a pair
of charged Higgs bosons ($H^\pm$).

The coupling constants in the Higgs potential of the MSSM are
determined
by the gauge couplings  due to the requirement of supersymmetry.
Hence, at the Born level, the masses of $H^\pm$ and $A$ in the MSSM
are related by the mass of the $W^\pm$ boson ($m_W$) as\footnote{
We note that in general, a CP
violating phase can enter the Higgs sector of the MSSM, so that
the CP-even Higgs bosons can mix with the CP-odd
Higgs scalar and the mass relation~(\ref{eq:massrel}) does
not hold any more. Here, we shall focus our study on the
MSSM with a CP invariant Higgs sector.
}
\begin{eqnarray}
M^2_{H^\pm} = M^2_A + m^2_W \, .
\label{eq:massrel}
\end{eqnarray}
Furtheemore, the coupling of
$W^-$-$A$-$H^+$ is induced from the gauge invariant
kinetic term of the Higgs sector \cite{hhg}:
\begin{eqnarray}
{\cal L}_{int}= \frac{g}{2} W^+_\mu (A \del^\mu H^- - H^- \del^\mu A)
+ {\rm h.c.} \, ,
\end{eqnarray}
and its strength is determined by the weak gauge coupling $g$.

Most production processes studied in the literature for testing the
MSSM contain at least two SUSY parameters (such as $\tan \beta$
and $M_A$) in the search for supersymmetric Higgs bosons.
Furthermore, the detection efficiency of the signal event depends on the
assumed decay channels of the SUSY particles, hence, on  the detailed
choice of SUSY parameters.
If the signal is not found after comparing experimental data with
theory prediction,
it is a common practice to constrain the product of
the production cross section and the decay branching ratios
of final state SUSY particles as a function of
the multiple-dimension SUSY parameter space of the MSSM.

In Ref.~\cite{wah_plb}, a novel proposal was made to study the
$AH^\pm$ production process at hadron colliders via 
\begin{equation}
q {\bar q'} \to W^{\pm\ast} \to A H^\pm \, .
\end{equation}
\begin{figure}[t]
\begin{center}
\scalebox{0.48}{\includegraphics{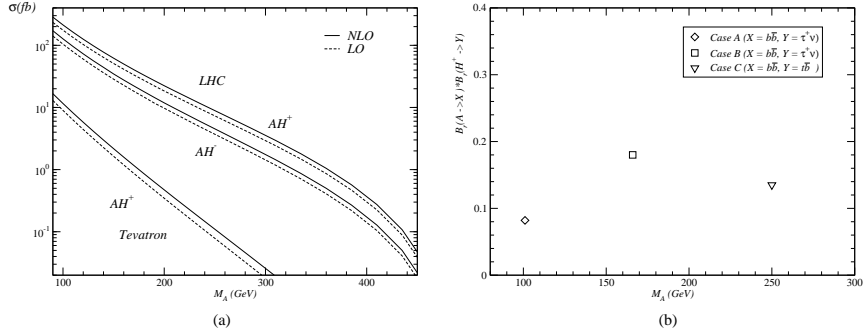}}
\end{center}
\caption{(a) The LO (dotted lines) and NLO QCD (solid lines)
cross sections of the $AH^+$ and $AH^-$ pairs
as a function of $M_A^{}$ at the Tevatron (a 1.96\,TeV $p \bar p$ collider),
and the LHC (a 14\,TeV $p p$ collider).
The cross sections for $AH^+$
and $AH^-$ pair productions coincide at the
Tevatron for being a $p \bar p$ collider.
(b) Constraints on the
product of branching ratios
$B(A \ra b {\bar b}) \times B(H^+ \ra \tau^+ \nu_\tau)$
as a function of $M_A$ for Case A and Case B, and
$B(A \ra b {\bar b}) \times B(H^+ \ra t {\bar b})$
for Case C, at the LHC, where
$\tau^+$ decays into $\pi^+ {\bar \nu}_\tau$ channel.
}
\label{fig:susy2003}
\vspace*{-0.3cm}
\end{figure}
In Fig.~\ref{fig:susy2003}(a), we show the inclusive production
rate of $A H^\pm$ as a function of
$M_A$ for the Tevatron (a 1.96\,TeV proton-antiproton collider)
and the LHC (a 14\,TeV proton-proton collider).
Typically, the next-to-leading order QCD production rate is about
20\% higher than the leading-order rate. The higher order ($\alpha^2_s$
or above) QCD correction is estimated to be about 10\% at
the Tevatron and less than a percent at the LHC, for $M_{A}^{}=120$
GeV, when the factorization scale is varied around
the $A H^\pm$ invariant mass $\sqrt{{s}}$ by a factor of 2.
The dominant one-loop electroweak corrections to the
$q{\bar q'} \to A H^\pm$ process come from the loops of top ($t$) and
bottom ($b$) quarks as well as
stops ($\tilde{t}_{1,2}$) and sbottoms ($\tilde{b}_{1,2}$)
due to their potentially large couplings to Higgs bosons.
However, as shown in Ref.~\cite{wah_prd}, the electroweak
radiative corrections are found to be at most a couple of
percent and smaller than the uncertainty
in higher order (beyond the NLO) QCD corrections,
parton distribution functions,
or the accuracy of the experimental measurement.

This process possesses the following interesting
properties.
\begin{itemize}
\item
  Its Born level production rate depends only on one
  SUSY parameter that can be determined by
  kinematic variables (e.g., the invariant mass of the
  $b \bar b $ pair from the decay of $A$).
\item
  Its higher order production rate is not sensitive to
  detailed SUSY parameters through radiative corrections.
\item
  Its final state particle kinematics can be properly modelled
  without specifying SUSY parameters.
  Hence, the detection efficiency of the signal
  can be accurately determined.
\item
  If the signal is found, it can be used to distinguish the MSSM
  from its alternatives such as the THDM, by testing
  the MSSM mass relation Eq.~(\ref{eq:massrel})
\item
  If the signal is not found, one can
  constrain the MSSM by limiting the
  product of decay branching ratios alone,
  without convoluting with the production cross section.
\end{itemize}

Either in the MSSM or the Type-II THDM,
for a large $\tan \beta$ value,
the dominant decay mode is $h,H,A \rightarrow b\ov{b}$
for the neutral Higgs bosons ($h,H,A$), and
$ H^+ \rightarrow \tau^+ \nu $
for 
$m_{H^\pm}^{} < m_t + m_b$.
Due to the missing energy carried away by the
final state neutrino,
it is not possible to directly reconstruct the mass of $H^+$
in the $\tau^+ \nu$ mode .
But, the transverse mass of 
$H^+$
can be reconstructed from the $\tau$ jet momentum
and missing transverse energy.
For $m_{H^{\pm}}\gtrsim 200$\,GeV, the dominant decay mode is
$H^{+}\rightarrow t \ov{b}$, in which $M_{H^+}$ can be reconstructed
after properly choosing the longitudinal momentum of the neutrino
(with a two-fold solution) from $t$ decay.

To test whether such a signal can be detected at the LHC, we
performed a Monte Carlo study at the parton level in 
Ref.~\cite{wah_prd}.
To cover both decay modes in our study, 
we consider the following three benchmark
cases with $\tan \beta = 40$:
\begin{itemize}
\item [(A)] $M_A=101$\,GeV (and  $M_{H^+} < m_t + m_b$), and
$H^{+}\rightarrow \tau \nu$ being the dominant decay mode.
\item [(B)] $M_A=166$\,GeV (and  $M_{H^+} \sim m_t + m_b$),
and $H^{+}\rightarrow \tau \nu$ being the dominant decay mode.
\item [(C)] $M_A=250$\,GeV and
$H^{+}\rightarrow t\ov{b}$ being the dominant decay mode.
\end{itemize}
We found that at the LHC this signal event can indeed provide
useful information about the MSSM Higgs sector.
If the $AH^+$ signal is not found in the decay mode of
$\tau^+ \ra \pi^+ {\bar \nu}_\tau$, then we can constrain the
product of branching ratios
$B(A \ra b {\bar b}) \times B(H^+ \ra \tau^+ \nu_\tau)$
as a function of $M_A$, as shown in Fig.~\ref{fig:susy2003}(b).
This corresponds to Case A or Case B.
In case C, for $M_{H^+} > m_t + m_b$,
not finding the signal event implies an upper bound on
 $B(A \ra b {\bar b}) \times B(H^+ \ra t {\bar b})$
for a given $M_A$.
Including the negatively charged channel $AH^-$ and the
$\rho \nu$  decay mode of $\tau$ can tighten the above bounds roughly
by a factor of $\sqrt{3}$. However, to have a more accurate
conclusion, a full event generator with detector simulation
should be used to repeat the analysis outlined in this paper.

From Fig.~\ref{fig:susy2003}(a), we see that the $AH^+$ rate becomes
very small (less than about $0.1$\,fb) at the LHC once $M_A$ is 
larger than 400\,GeV. Hence,
to cover the whole mass spectrum of the TeV scale MSSM, we
need a high energy collider that can be sensitive to this process
for $M_A$ approaching the TeV region.
This could be one of the motivations for proposing a future
Very Large Hadron Collider (VLHC), a 200\,TeV proton-proton collider.


\end{document}